\documentclass[sigconf]{acmart}

\hypersetup{draft}
\usepackage{amsmath,amssymb,amsfonts}
\usepackage{algorithmic}
\usepackage{graphicx}
\usepackage{textcomp}
\usepackage{xcolor}
\usepackage{booktabs}
\usepackage{enumitem}
\setlist[itemize]{leftmargin=*}

\setcopyright{rightsretained}

\begin{document}

\newcommand{\sstub}{SStuB}
\newcommand{\sstubs}{SStuBs}
\newcommand{\manysstubs}{Many\sstubs4J}

\title{How Often Do Single-Statement Bugs Occur?\\ The \manysstubs\ Dataset
}

\author{Rafael-Michael Karampatsis}
 \affiliation{
   \institution{University of Edinburgh}
   \city{Edinburgh}
   \country{United Kingdom}
}
\author{Charles Sutton}
 \affiliation{
   \institution{Google Research, University of Edinburgh and The Alan Turing Institute}
   \city{Mountain View, CA}
   \country{United States}
}

\thanks{This work was supported in part by the EPSRC Centre for Doctoral Training in Data Science, funded by the UK Engineering and Physical Sciences Research Council (grant EP/L016427/1) and the University of Edinburgh.
}

\begin{abstract}
  Program repair is an important but difficult software engineering problem.
  One way to achieve acceptable performance %
  is to focus on classes of simple bugs, such as bugs with single statement fixes,
  or that match a small set of bug templates.
However, it is very difficult to estimate the recall of repair techniques for simple bugs,
as there are no datasets about how often the associated bugs occur in code.
To fill this gap, we provide a dataset of  153,652 single statement bug-fix changes mined from 1,000 popular open-source Java projects,
annotated by whether they match any of a set of 16 bug templates, inspired by state-of-the-art program repair techniques.
In an initial analysis, we find that about 33\% of the simple bug fixes match the templates, indicating that a remarkable number of single-statement bugs
can be repaired with a relatively small set of templates.
Further, we find that template fitting bugs appear with a frequency of about one bug per 1,600-2,500 lines of code (as measured by the size of the project's latest version).
We hope that the dataset will prove a resource for both future work in program repair and studies in empirical software engineering.
\end{abstract}


\copyrightyear{2020}
\acmYear{2020}
\acmConference[MSR '20]{17th International Conference on Mining Software Repositories}{October 5--6, 2020}{Seoul, Republic of Korea}
\acmBooktitle{17th International Conference on Mining Software Repositories (MSR '20), October 5--6, 2020, Seoul, Republic of Korea}\acmDOI{10.1145/3379597.3387491}
\acmISBN{978-1-4503-7517-7/20/05}

\maketitle

\section{Introduction}

Fixing bugs in programs, that is, program repair, is one of the core tasks in software maintenance,
but requires effort to
 analyze failed executions, locate the cause of the fault, synthesize a bug fix and validate that the fault has been corrected without introducing new ones \cite{Mullerburg1983}.
 Automatic program repair \cite{LeGoues2012, Long2016, Pradel2018, Monperrus:survey} attempts to alleviate most of the manual effort of locating and repairing faults.
However, a major concern in industry is that linters and program repair methods approaches are required to have high precision without risking achieving high enough recall.
As an industrial example Google's Tricorder \cite{Sadowski2015} enforces a false positive rate $< 10\%$.

One way to find a ``sweet spot'' of maintaining high precision with adequate recall is to focus on
repairing types of simple bugs, such as one-line bugs,
or  bugs that fall into a small set of templates,
such as mutation operators
\cite{LeGoues2012} or other types of predefined templates \cite{Long2015, Long2016, Pradel2018}.
However, these have been evaluated on either a relatively small numbers of projects, e.g. 69 defects in 8 applications or on synthetic data.
Because of this lack of data, it has not previously been possible to estimate the \emph{recall} of a set of repair templates,
that is, the percentage of real-world bugs that can be repaired by one of the templates.
Simultaneously to the current work, a larger dataset of one-line bugs has been mined \cite{Sequencer},
but even this dataset does not attempt to classify bugs into templates.

Aiming to fill this gap, we provide a dataset containing 25,539 single-statement bug-fix changes mined from 100 popular open-source Java Maven projects as well as a larger one containing 153,652 single-statement bug-fix changes mined from 1,000 popular open-source Java projects,
annotated by whether they match any of a set of 16 bug templates, inspired by state-of-the-art program repair techniques.
The chosen templates aim at extracting bugs that compile both before and after repair as such can be quite tedious to manually spot, yet their fixes are so simple that many developers would call them ``stupid'' upon realization.
We will refer onwards to these bugs as ``simple stupid bugs'' (\sstubs)\footnote{The
acronym is intended to reflect the fact that, for the authors at least, finding such a bug can feel much like stubbing one's toe.} and the corresponding dataset as the \manysstubs\ dataset. %
Automatic repair of \sstubs\ is potentially an intermediate step toward more general program repair tools,
while already being useful to developers.
We also think that \sstubs\ might be a good start for the evaluation of machine learning based fault localization and repair methods.

An extra distinctive feature of our dataset is that the smaller version is restricted to projects that can be built automatically
using Maven.
Those that contain a test suite can be built and used to evaluate test based techniques.
In an initial analysis, we find that 33.04\% in the smaller version dataset and 33.47\% in the larger version of all of the single-statement bugs that we mine match at least one of the \sstub\ templates resulting in 10,231 and 63,923 \sstub\ instances respectively.
This indicates that a remarkable number of singe-statement bugs can be repaired with a relatively small set of templates.
In further analysis we also estimated the frequency in lines of code with which these pattern based and general single-statement bugs appear.
This estimation is based on the size of the project's latest version and reveals that in the smaller dataset version SStuBs appear with a frequency of about 1 per 1,600 lines of code and 1 per 2,500 lines of code for the large version.
We hope that this dataset can serve as a valuable resource for both future work in program repair and studies in empirical software engineering.

\section{Methodology}
We next describe the methodology we employed to build the dataset.
Our data generation tools along with documentation and detailed instructions for how to use them
are available in a public GitHub repository\footnote{https://github.com/mast-group/mineSStuBs} 
and the dataset is publicly available in Zenodo.\footnote{DOI: https://doi.org/10.5281/zenodo.3653444}

\subsection{Selecting Appropriate Java Projects}
In order to mine a high quality dataset we opted to selecting high popularity projects.
For the small version of the dataset we selected the 100 most popular open source Java Maven \cite{Miller2010} projects from GitHub up to 1/4/2017. 
To allow evaluation of repair tools that might require building the projects, we selected only Maven ones because
it is easy to automatically download the required dependencies for every project and build it.
In contrast, manual downloading of dependencies would require an immense amount of human effort.
To create a ranking for the projects we downloaded the MySQL dump of GHTorrent \cite{Gousios2013} up to 1/4/2017. 
A project's popularity is determined by computing the sum of z-scores of its forks and stars \cite{Allamanis2015a, Allamanis2016}.
Lastly, we pulled the projects' head commit by 28/1/2019 and considered commits until that date.
The same approach was used to rank projects for the larger version.
However, the ranking was calculated using a later dump of GHTorrent from 1/1/2019.
A download script along with the list of projects for both variants of the dataset are also available to ensure replicability.

\subsection{Classifying Commits as Bug-Fixing or not}
For every project our tool searches historically through all of its commits to locate bug-fixing ones. 
To decide if a commit fixes a bug, we checked if its
commit message contains at least one of the
keywords: `error', `bug', `fix', `issue', `mistake', `incorrect', `fault', `defect', `flaw',
and `type'. 
This heuristic was previously used by Ray et al. \cite{Ray2015} and was shown to achieve 96\% accuracy on a set of 300 manually verified commits and 97.6\% on a set of 384 manually verified commits \cite{Tufano2018}.
We sampled 100 random commits containing \sstubs\ from the small version of the dataset and found it to achieve 94\% accuracy.
The above process produced a total of 115,929 and 883,982 bug-fixing  commits for the small and large dataset variants.

\subsection{Selecting Single Statement Changes}
We have opted to restrict the dataset to small bug fixes that do not require much code modification to fix.
Additionally, we are interested in bugs that are not just syntactic errors but cases where the code compiles both before and after the bug was located and repaired.
As we are interested in simple bugs that involve only a single statement, we filter out any commits that either add or delete a Java file.
We also filter out commits which make a multiple-statement change at any single position in the Java file.
We do \emph{not} filter out commits that make single-line modifications at more than one position in the same file.
Similarly to the diff algorithm, we consider a modification as deleting the old lines/statements and then adding the new ones.
To estimate whether a modification spans across multiple statements we calculate the diff for each modified Java file,
and 
for each modified chunk, we count how many statements were modified.
In the case of blocks each statement in the block's body is counted as a different statement. 
For \texttt{if} and \texttt{while} statements, we count the condition as a separate statement
for this purpose.
This method allows to us include fixes to single simple statements that span across multiple lines (e.g. due to stylistic reasons)
as a simple fix, unlike a line-based approach.
Any commits that modify multiple statements in any single position returned by the diff are dropped while we still maintain commits for which a file's diff contains multiple positions with single statement modifications.
In the first case it is not trivial to align the deleted and added statements while it is in the latter.
For example, one or more of the deleted statements may have been replaced by multiple of the added ones while simultaneously one or more of the deleted statements may have simply been deleted.
We note that our tool ignores any changes to comments, blank lines as well as any formatting changes.
Our methodology allows cases where the same expression containing a bug appeared multiple times in the file.
This filtering produces almost 13,000 and 86,769 commits for the two dataset versions.
Lastly, the employed methodology works in a similar way to the popular SZZ algorithm \cite{Sliwerski2005} and its extensions \cite{Kim2006, Williams2008} that have extensively been used to spot fix inducing changes.

\subsection{Creating Abstract Syntax Trees}

Each file in the commit that contains one or more bugs is parsed, yielding an abstract syntax tree (AST) of the file before the repair.
Then, for each repaired line in the file we extract the AST after applying the repair only on that line and leaving the rest of the lines as is.
Each extracted pair of ASTs (original and single fix) only differ on the node(s) for the modified line.
By performing a simultaneous depth-first traversal on the two ASTs we locate the first node on which the two ASTs differ.

\subsection{Filtering out Clear Refactorings}

Although we filter for bug-fixing changes in Step B,
there might still exist changes in the data that do not fix a bug or that do not even produce any behavioural changes.
This could happen because the commit-message filter had
a false positive, or because the change is tangled \cite{herzig2013tangled}, and contains a bug-fixing modification
along with unrelated ones to other files.
To reduce the number of non-fixing changes in the dataset, we observe that
there is a class of refactorings that can produce small changes, namely renamings.
These are extracted via the diffs of the modified files.
Our method spots variable, function, or class renaming as well as any uses of them across other modified files in the commit and excludes them.

\subsection{\sstub\ Patterns}

We next describe the 16 \sstub\ patterns.
We opted to choose patterns that appear often.
Many of these have been used in pattern-based repair and mutation tools \cite{LeGoues2012, Long2015, Long2016, Pradel2018}.
Here we provide a brief description of each pattern. 
Due to page limitations we do not include examples here but in the README of the GitHub repository. %
\begin{itemize}
  
\item \emph{Change Identifier Used}
Checks whether an identifier appearing in some expression in the statement was replaced with another one.
It is easy for developers to by accident utilize a different identifier than the intended one that has the same type.
Copy pasting code is a potential source of such errors.
Similarly named identifiers may further contribute to the occurrence of such errors.

\item \emph{Change Numeric Literal}
Checks whether a numeric literal was replaced with another one.
It is easy for developers to mix two numeric values in their program.

\item \emph{Change Boolean Literal}
Checks whether a Boolean literal was replaced. True is replaced with False and vice-versa.
In many cases developers use the opposite Boolean value than the intended one.

\item \emph{Change Modifier}
Checks whether a variable, function, or class was declared with the wrong modifiers.
For example a developer can forget to declare one of the modifiers.

\item \emph{Wrong Function Name}
Checks if a function with the same parameter list but the wrong name was called. This is a usual pitfall.

\item \emph{Same Function More Args}
Checks whether an overloaded version of the function with more arguments was called.
Functions with multiple overload can often confuse developers.

\item \emph{Same Function Less Args}
Checks whether an overloaded version of the function with less arguments was called.
For instance, a developer can forget to specify one of the arguments and not realize it if the code still compiles due to function overloading.

\item \emph{Same Function Change Caller}
Checks whether in a function call expression the caller object for it was replaced with another one.
When there are multiple variables with the same type a developer can accidentally perform an operation.
Copy pasting code or mixing similar variables are common cases of such errors.

\item \emph{Same Function Swap Args}
Checks whether a function was called with two of its arguments swapped.
When multiple function arguments are of the same type, developers %
can easily swap two of them without realizing.
It was also used in DeepBugs \cite{Pradel2018}. %

\item \emph{Change Binary Operator}
Checks whether a binary operand was accidentally replaced with another one of the same type.
For example, developers very often mix comparison operators in expressions.
A similar pattern was also used in DeepBugs \cite{Pradel2018}.

\item \emph{Change Unary Operator}
Checks whether a unary operand was accidentally replaced with another one of the same type
(e.g., developers often forget the ! operator in a boolean expression).

\item \emph{Change Operand}
Checks whether one of the operands in a binary operation was wrong.
It was also used in DeepBugs \cite{Pradel2018}.

\item \emph{More Specific If}
Checks whether an extra condition (\&\& operand) was added  in an \texttt{if} statement's condition.

\item \emph{Less Specific If}
Checks whether an extra condition which either this or the original one needs to hold ($\|$ operand) was added in an \texttt{if} statement's condition.

\item \emph{Missing Throws Exception}
Checks whether the fix added a \texttt{throws} clause in a function declaration.

\item \emph{Delete Throws Exception}
Checks whether the fix deleted a \texttt{throws} clause in a function declaration.
\end{itemize}

\subsection{\sstub\ Pattern Matching}

Finally, each pair of ASTs is automatically checked for fitting any of the \sstub\ patterns.
Each pattern is expressed as a mutation operation on the original AST that produces the new one.
All instances are added to the single-statement dataset, while only those that match \sstub\ patterns are saved in the \sstubs\ one.

\section{\manysstubs\ Dataset Statistics}

\begin{table}[tpb]
\renewcommand{\arraystretch}{1.05}
\caption{Statistics for each SStuB pattern.}
\begin{center}
\resizebox{0.935\columnwidth}{!}{
\begin{tabular}{l r r r r}
\toprule
\textbf{Pattern Name} & \textbf{\textit{SStuBs}} & \textbf{\textit{Ratio}} & \textbf{\textit{SStuBs L}} & \textbf{\textit{Ratio L}} \\
\midrule
Change Identifier Used 		& 3265 	& 12.78\%	& 22668	& 14.75\% \\
Change Numeric Literal 		& 1137 	& 4.45\%	& 5447	& 3.55\% \\
Change Modifier 			& 1852 	& 7.25\%	& 5011	& 3.26\% \\
Change Boolean Literal 		& 169 	& 0.66\%	& 1842	& 1.20\% \\
\midrule
Wrong Function Name 		& 1486 	& 5.82\%	& 10179	& 6.62\% \\
Same Function More Args 	& 758 	& 2.97\%	& 5100	& 3.32\% \\
Same Function Less Args 	& 179 	& 0.70\%	& 1588	& 1.03\% \\
Same Function Wrong Caller	& 187 	& 0.73\%	& 1504	& 0.98\% \\
Same Function Swap Args 	& 127 	& 0.50\%	& 612	& 0.39\% \\
\midrule
Change Binary Operator 		& 275 	& 1.08\%	& 2241	& 1.46\% \\
Change Unary Operator 		& 170 	& 0.67\%	& 1016	& 0.66\% \\
Change Operand 				& 120 	& 0.47\%	& 807	& 0.53\% \\
\midrule
Less Specific If 			& 215 	& 0.84\%	& 2813	& 1.83\% \\
More Specific If			& 175 	& 0.69\%	& 2381	& 1.55\% \\
Missing Throws Exception 	& 68 	& 0.27\%	& 206	& 0.13\% \\
Delete Throws Exception 	& 48 	& 0.19\%	& 508	& 0.33\% \\
\midrule
TOTAL NO DOUBLE COUNTS		& 8438	& 33.04\%	& 51433	& 33.47\% \\
TOTAL 						& 10231	& 40.06\%	& 63923	& 41.59\% \\
\bottomrule
\end{tabular}
}
\label{tab:sstubPercentages}
\end{center}
\end{table}

The \manysstubs\ dataset consists of 10,231 and 63,923 instances of single statement bugs mined from 12,598 and 86,771 bug-fix commits with only single-statement changes respectively for each version.
Consequently, on average almost 2 single statement bugs and 0.75 \sstubs\ were mined per valid commit.
The data is saved in JSON files and detailed information is available in the GitHub repository.
Each \sstub\ instance is also annotated with the \sstub\ pattern satisfied, the project's name, the Java file's name, the hashes of the fix inducing commit and its parent, the line at which the bug starts, and the AST subtree's location. 
In some cases a statement might fit more than one patterns.
In those cases it is counted as separate instances.
However, in most cases the patterns are distinct.
The statistics for each of the 16 \sstub\ patterns of the \manysstubs\ dataset are shown in Table~\ref{tab:sstubPercentages}.
Patterns that are similar are grouped together (e.g. patterns that concern functions) and sorted in descending frequency order.
The three most common \sstub\ patterns are \emph{Change Identifier Used}, \emph{Wrong Function Name}, and \emph{Change Numeric Literal}.

We note that the mined bugs have not been annotated by severity and we expect that to vary.
Some of the bugs appear in test code.
Although bugs in test code will not reach a final product, they can have significant effect on it as they can potentially mask important bugs in it.
Test oracle errors can bring confusion that slows down the debugging process while fixing them improves the performance of fault localization algorithms \cite{Guo2015}.
Such bugs might also be quite tedious to locate as it is very rare to test a test suite and even if we follow that logic we would have to endlessly create tests for the tests.
By design we do not attempt to restrict the bugs to those that have a failing test case.
The goal is to reproduce the situtations that the bugs happen in the wild.
Lastly, as it was recently shown, unit tested code does not appear to be associated with fewer failures while increased coverage is associated with more failures \cite{chioteli2019does}.

\section{Research Questions} \label{sec:questions}

Although the paper focuses on the dataset, we run a simple analysis to support
our design decision to focus our new dataset on \sstubs.
In order to explore whether the \sstub\ patterns are useful targets for program repair techniques,
we asked two research questions.

\subsection*{RQ1. Are \sstubs\ common in open-source code?}%
\label{sec:questionPercentage}
We measured for each SStuB type the percentage of single statement modifications that are not clear refactorings and fit the pattern.
These are visualized in Table~\ref{tab:sstubPercentages}.
For each project $P$ we also estimated the following two densities for the mined SStuBs:
(a) the number of \sstubs\ in project $P$ / total lines in $P$ at the final snapshot and
(b) the number of \sstubs\ in project $P$ / total lines added and deleted in $P$ by the final snapshot. 
Thus, estimating the frequency per line of code modifications in the project's history. That is counting any line that was added or deleted to the project from the start to its latest version.
A line modification is counted twice (once as a deletion and once as an addition). 
Once for deleting the old and once for adding the new line.
Comments and empty lines were excluded from these estimations.
We found that in the smaller version of the dataset SStuBs appear with densities of about 2,400 and 30,000 lines of code (LOC) respectively.

We also estimated the same densities for the larger dataset variant.
We found that such bugs appear with a frequency of about 1,600 and 20,000 LOC respectively.
As a threat to validity, we acknowledge that the number of LOC in the final snapshot may not be the most informative denominator for a measure of bug density, but developing better ones is a thorny issue left for future work.

\subsection*{RQ2. Can SStuBs be spotted by existing tools such as static analyzers?} \label{questionSpot}

We measure the proportion of bugs in our dataset that can be identified by
the popular static analysis tool SpotBugs.\footnote{https://spotbugs.github.io/}
If SpotBugs reports any bug for the line containing the \sstub\ then we consider that SpotBugs successfully detected it.
We find that SpotBugs could only locate about $12\%$ of SStuBs while also reporting more than $200$ million possible bugs when configured to report all warnings, even those with low confidence.
In fact, as explained the actual recall is even lower.
This is confirmed by a recent study where three static bug detectors including SpotBugs located only $4.5\%$ of bugs \cite{Habib2018}.
This means that a developer would have to look through hundreds of thousands of warnings produced by SpotBugs to locate a single SStuB.
This highlights the necessity for tools that are specifically built to detect \sstubs.
The scripts used to run and evaluate SpotBugs are also available in our repository.

\section{Related Work}

Several previous data sets of real-world bugs have been curated.
Defects4J \cite{Just2014} is a popular dataset consisting 395 Java bugs.
Each bug is fixed in a single commit but the fix may modify multiple source code lines.
The ManyBugs dataset \cite{Goues2015} contains 185 C bugs, a subset of which were used by the GenProg \cite{LeGoues2012}, Prophet \cite{Long2016} and SPR \cite{Long2015} papers.
Bugs.jar \cite{Saha2018} is comprised of 1,158 Java bugs and their patches.
These datasets have the disadvantage of being relatively small.
More recently, a few larger-scale data sets of small bugs have been created.
The combined datasets are the CodRep dataset \cite{CodRep} and the Bugs2Fix dataset \cite{Bugs2Fix} resulting in 40,289 one-line bugs.
These datasets are combined into a single dataset of one line bugs in \cite{Sequencer}.
Our datasets are of similar size consisting of 25,539 and 153,652 single-statement bugs.
In contrast, our dataset focus on estimating the frequency of \sstub\ templates, motivated by recent
program repair tools and also operates on the statement level, which prevents falsely excluding instances due to formatting or stylistic reasons.
Also, the projects from which the small version of our dataset was generated can easily be built using Maven and we provide a list of projects containing tests and which tests fail for each instance (in GitHub repo).
Thus test based methods can be evaluated upon them.
However, unlike Defects4J that aims in comparing test-based patch generation approaches, it aims in techniques that can accurately highlight SStuBs early in development allowing immediate patching since in many cases the fix might be trivial.
Lastly, unlike previous datasets, we take additional steps to filter out refactorings, although we acknowledge that such instances might be rare.
In our case however, we were able to filter out almost 5,000 and 35,000 refactored statements for the two dataset versions.

\section{Limitations - Threats to Validity}

Although unlikely, it is possible for our SZZ like methodology to extract a pair of aligned statements that are unrelated (i.e. one line was deleted and one was added).
We do spot refactorings but there is no guarantee that we have detected 100\% of them.
The heuristic used to spot bug fixing commits could introduce false positives, but this is mitigated by the fact that we focus on single line commits and as already discussed the false positive rate is low.
Our dataset will not be useful for evaluating whether repair systems are good at fixing larger bugs.
Our dataset is restricted to Java but could be replicated for other languages by using a parser and creating a module that checks if an AST pair fits any of the SStuB patterns.
The precise set of patterns might vary across languages and determining these might be an interesting direction for future work.

\section{Conclusions}

We introduce a new, large-scale dataset of real-world \sstubs, simple one-statement bugs,
in Java for the evaluation of program repair techniques. The distinguishing feature of our dataset is
that where possible, the \sstubs\ are categorized into one of 16 bug templates, which are inspired
by those considered in state-of-the-art program repair methods.
These types of bugs often result in code that compiles, which means that they are particularly interesting for automated repair. 
We find that \sstubs\ occur relatively often --- one per 1,600 LOC in the projects
we study --- making them potentially a promising evaluation dataset for repair techniques that could be used to estimate their actual recall.
The data could also be used to answer other research questions, such as empirical questions about how and when simple
bugs are introduced, or about evaluating program repair techniques for small bugs.
Also, it can aid in evaluating machine learning systems that learn to localize simple bugs via examples \cite{Pradel2018} or a language model's entropy \cite{Karampatsis2020big}.
Last, coverage information for the maven projects with tests suits in the dataset could be used to estimate how often do tests cover SStuBs.

\bibliographystyle{ACM-Reference-Format}
\bibliography{sstubs}


\begin{thebibliography}{27}


\ifx \showCODEN    \undefined \def \showCODEN     #1{\unskip}     \fi
\ifx \showDOI      \undefined \def \showDOI       #1{#1}\fi
\ifx \showISBNx    \undefined \def \showISBNx     #1{\unskip}     \fi
\ifx \showISBNxiii \undefined \def \showISBNxiii  #1{\unskip}     \fi
\ifx \showISSN     \undefined \def \showISSN      #1{\unskip}     \fi
\ifx \showLCCN     \undefined \def \showLCCN      #1{\unskip}     \fi
\ifx \shownote     \undefined \def \shownote      #1{#1}          \fi
\ifx \showarticletitle \undefined \def \showarticletitle #1{#1}   \fi
\ifx \showURL      \undefined \def \showURL       {\relax}        \fi
\providecommand\bibfield[2]{#2}
\providecommand\bibinfo[2]{#2}
\providecommand\natexlab[1]{#1}
\providecommand\showeprint[2][]{arXiv:#2}

\bibitem[\protect\citeauthoryear{Allamanis, Barr, Bird, and Sutton}{Allamanis
  et~al\mbox{.}}{2015}]%
        {Allamanis2015a}
\bibfield{author}{\bibinfo{person}{Miltiadis Allamanis},
  \bibinfo{person}{Earl~T. Barr}, \bibinfo{person}{Christian Bird}, {and}
  \bibinfo{person}{Charles Sutton}.} \bibinfo{year}{2015}\natexlab{}.
\newblock \showarticletitle{Suggesting Accurate Method and Class Names}. In
  \bibinfo{booktitle}{\emph{Proceedings of the 2015 10th Joint Meeting on
  Foundations of Software Engineering}} (Bergamo, Italy)
  \emph{(\bibinfo{series}{ESEC/FSE 2015})}. \bibinfo{publisher}{ACM},
  \bibinfo{address}{New York, NY, USA}, \bibinfo{pages}{38--49}.
\newblock
\showISBNx{978-1-4503-3675-8}
\urldef\tempurl%
\url{https://doi.org/10.1145/2786805.2786849}
\showDOI{\tempurl}


\bibitem[\protect\citeauthoryear{Allamanis, Peng, and Sutton}{Allamanis
  et~al\mbox{.}}{2016}]%
        {Allamanis2016}
\bibfield{author}{\bibinfo{person}{Miltiadis Allamanis}, \bibinfo{person}{Hao
  Peng}, {and} \bibinfo{person}{Charles~A. Sutton}.}
  \bibinfo{year}{2016}\natexlab{}.
\newblock \showarticletitle{A Convolutional Attention Network for Extreme
  Summarization of Source Code}. In \bibinfo{booktitle}{\emph{Proceedings of
  {ICML} 2016}}, Vol.~\bibinfo{volume}{48}. \bibinfo{pages}{2091--2100}.
\newblock
\urldef\tempurl%
\url{http://proceedings.mlr.press/v48/allamanis16.html}
\showURL{%
\tempurl}


\bibitem[\protect\citeauthoryear{Chen, Kommrusch, Tufano, Pouchet, Poshyvanyk,
  and Monperrus}{Chen et~al\mbox{.}}{2019}]%
        {Sequencer}
\bibfield{author}{\bibinfo{person}{Zimin Chen}, \bibinfo{person}{Steve
  Kommrusch}, \bibinfo{person}{Michele Tufano},
  \bibinfo{person}{Louis{-}No{\"{e}}l Pouchet}, \bibinfo{person}{Denys
  Poshyvanyk}, {and} \bibinfo{person}{Martin Monperrus}.}
  \bibinfo{year}{2019}\natexlab{}.
\newblock \showarticletitle{SequenceR: Sequence-to-Sequence Learning for
  End-to-End Program Repair}.
\newblock \bibinfo{journal}{\emph{CoRR}}  \bibinfo{volume}{abs/1901.01808}
  (\bibinfo{year}{2019}).
\newblock
\showeprint[arxiv]{1901.01808}
\urldef\tempurl%
\url{http://arxiv.org/abs/1901.01808}
\showURL{%
\tempurl}


\bibitem[\protect\citeauthoryear{Chen and Monperrus}{Chen and
  Monperrus}{2018}]%
        {CodRep}
\bibfield{author}{\bibinfo{person}{Zimin Chen} {and} \bibinfo{person}{Martin
  Monperrus}.} \bibinfo{year}{2018}\natexlab{}.
\newblock \showarticletitle{The CodRep Machine Learning on Source Code
  Competition}.
\newblock \bibinfo{journal}{\emph{CoRR}}  \bibinfo{volume}{abs/1807.03200}
  (\bibinfo{year}{2018}).
\newblock
\showeprint[arxiv]{1807.03200}
\urldef\tempurl%
\url{http://arxiv.org/abs/1807.03200}
\showURL{%
\tempurl}


\bibitem[\protect\citeauthoryear{Chioteli, Batas, and Spinellis}{Chioteli
  et~al\mbox{.}}{2019}]%
        {chioteli2019does}
\bibfield{author}{\bibinfo{person}{Efstathia Chioteli},
  \bibinfo{person}{Ioannis Batas}, {and} \bibinfo{person}{Diomidis Spinellis}.}
  \bibinfo{year}{2019}\natexlab{}.
\newblock \showarticletitle{Does Unit-Tested Code Crash? A Case Study of
  Eclipse}.
\newblock \bibinfo{journal}{\emph{arXiv preprint arXiv:1903.04055}}
  (\bibinfo{year}{2019}).
\newblock


\bibitem[\protect\citeauthoryear{Goues, Holtschulte, Smith, Brun, Devanbu,
  Forrest, and Weimer}{Goues et~al\mbox{.}}{2015}]%
        {Goues2015}
\bibfield{author}{\bibinfo{person}{C.~Le Goues}, \bibinfo{person}{N.
  Holtschulte}, \bibinfo{person}{E.~K. Smith}, \bibinfo{person}{Y. Brun},
  \bibinfo{person}{P. Devanbu}, \bibinfo{person}{S. Forrest}, {and}
  \bibinfo{person}{W. Weimer}.} \bibinfo{year}{2015}\natexlab{}.
\newblock \showarticletitle{The ManyBugs and IntroClass Benchmarks for
  Automated Repair of C Programs}.
\newblock \bibinfo{journal}{\emph{IEEE Transactions on Software Engineering}}
  \bibinfo{volume}{41}, \bibinfo{number}{12} (\bibinfo{date}{Dec}
  \bibinfo{year}{2015}), \bibinfo{pages}{1236--1256}.
\newblock
\showISSN{0098-5589}
\urldef\tempurl%
\url{https://doi.org/10.1109/TSE.2015.2454513}
\showDOI{\tempurl}


\bibitem[\protect\citeauthoryear{Gousios}{Gousios}{2013}]%
        {Gousios2013}
\bibfield{author}{\bibinfo{person}{Georgios Gousios}.}
  \bibinfo{year}{2013}\natexlab{}.
\newblock \showarticletitle{The GHTorrent dataset and tool suite}. In
  \bibinfo{booktitle}{\emph{Proceedings of the 10th Working Conference on
  Mining Software Repositories}} (San Francisco, CA, USA)
  \emph{(\bibinfo{series}{MSR '13})}. \bibinfo{publisher}{IEEE Press},
  \bibinfo{address}{Piscataway, NJ, USA}, \bibinfo{pages}{233--236}.
\newblock
\showISBNx{978-1-4673-2936-1}
\urldef\tempurl%
\url{http://dl.acm.org/citation.cfm?id=2487085.2487132}
\showURL{%
\tempurl}


\bibitem[\protect\citeauthoryear{{Guo}, {Zhou}, {Song}, {Gu}, and {Sun}}{{Guo}
  et~al\mbox{.}}{2015}]%
        {Guo2015}
\bibfield{author}{\bibinfo{person}{X. {Guo}}, \bibinfo{person}{M. {Zhou}},
  \bibinfo{person}{X. {Song}}, \bibinfo{person}{M. {Gu}}, {and}
  \bibinfo{person}{J. {Sun}}.} \bibinfo{year}{2015}\natexlab{}.
\newblock \showarticletitle{First, Debug the Test Oracle}.
\newblock \bibinfo{journal}{\emph{IEEE Transactions on Software Engineering}}
  \bibinfo{volume}{41}, \bibinfo{number}{10} (\bibinfo{date}{Oct}
  \bibinfo{year}{2015}), \bibinfo{pages}{986--1000}.
\newblock
\showISSN{2326-3881}
\urldef\tempurl%
\url{https://doi.org/10.1109/TSE.2015.2425392}
\showDOI{\tempurl}


\bibitem[\protect\citeauthoryear{Habib and Pradel}{Habib and Pradel}{2018}]%
        {Habib2018}
\bibfield{author}{\bibinfo{person}{Andrew Habib} {and} \bibinfo{person}{Michael
  Pradel}.} \bibinfo{year}{2018}\natexlab{}.
\newblock \showarticletitle{How Many of All Bugs Do We Find? A Study of Static
  Bug Detectors}. In \bibinfo{booktitle}{\emph{Proceedings of the 33rd ACM/IEEE
  International Conference on Automated Software Engineering}} (Montpellier,
  France) \emph{(\bibinfo{series}{ASE 2018})}. \bibinfo{publisher}{Association
  for Computing Machinery}, \bibinfo{address}{New York, NY, USA},
  \bibinfo{pages}{317–328}.
\newblock
\showISBNx{9781450359375}
\urldef\tempurl%
\url{https://doi.org/10.1145/3238147.3238213}
\showDOI{\tempurl}


\bibitem[\protect\citeauthoryear{Herzig and Zeller}{Herzig and Zeller}{2013}]%
        {herzig2013tangled}
\bibfield{author}{\bibinfo{person}{Kim Herzig} {and} \bibinfo{person}{Andreas
  Zeller}.} \bibinfo{year}{2013}\natexlab{}.
\newblock \showarticletitle{The impact of tangled code changes}. In
  \bibinfo{booktitle}{\emph{Working Conference on Mining Software
  Repositories}}. IEEE Press, \bibinfo{pages}{121--130}.
\newblock


\bibitem[\protect\citeauthoryear{Just, Jalali, and Ernst}{Just
  et~al\mbox{.}}{2014}]%
        {Just2014}
\bibfield{author}{\bibinfo{person}{Ren{\'e} Just}, \bibinfo{person}{Darioush
  Jalali}, {and} \bibinfo{person}{Michael~D. Ernst}.}
  \bibinfo{year}{2014}\natexlab{}.
\newblock \showarticletitle{Defects4J: A Database of Existing Faults to Enable
  Controlled Testing Studies for Java Programs}. In
  \bibinfo{booktitle}{\emph{Proceedings of the 2014 International Symposium on
  Software Testing and Analysis}} (San Jose, CA, USA)
  \emph{(\bibinfo{series}{ISSTA 2014})}. \bibinfo{publisher}{ACM},
  \bibinfo{address}{New York, NY, USA}, \bibinfo{pages}{437--440}.
\newblock
\showISBNx{978-1-4503-2645-2}
\urldef\tempurl%
\url{https://doi.org/10.1145/2610384.2628055}
\showDOI{\tempurl}


\bibitem[\protect\citeauthoryear{Karampatsis, Babii, Robbes, Sutton, and
  Janes}{Karampatsis et~al\mbox{.}}{2020}]%
        {Karampatsis2020big}
\bibfield{author}{\bibinfo{person}{Rafael-Michael Karampatsis},
  \bibinfo{person}{Hlib Babii}, \bibinfo{person}{Romain Robbes},
  \bibinfo{person}{Charles Sutton}, {and} \bibinfo{person}{Andrea Janes}.}
  \bibinfo{year}{2020}\natexlab{}.
\newblock \bibinfo{title}{Big Code != Big Vocabulary: Open-Vocabulary Models
  for Source Code}.
\newblock
\newblock
\showeprint[arxiv]{cs.SE/2003.07914}


\bibitem[\protect\citeauthoryear{Kim, Zimmermann, Pan, and Whitehead}{Kim
  et~al\mbox{.}}{2006}]%
        {Kim2006}
\bibfield{author}{\bibinfo{person}{S. Kim}, \bibinfo{person}{T. Zimmermann},
  \bibinfo{person}{K. Pan}, {and} \bibinfo{person}{E.~J.~Jr. Whitehead}.}
  \bibinfo{year}{2006}\natexlab{}.
\newblock \showarticletitle{Automatic Identification of Bug-Introducing
  Changes}. In \bibinfo{booktitle}{\emph{21st IEEE/ACM International Conference
  on Automated Software Engineering (ASE'06)}}. \bibinfo{pages}{81--90}.
\newblock
\showISSN{1938-4300}
\urldef\tempurl%
\url{https://doi.org/10.1109/ASE.2006.23}
\showDOI{\tempurl}


\bibitem[\protect\citeauthoryear{{Le Goues}, Nguyen, Forrest, and Weimer}{{Le
  Goues} et~al\mbox{.}}{2012}]%
        {LeGoues2012}
\bibfield{author}{\bibinfo{person}{Claire {Le Goues}}, \bibinfo{person}{ThanhVu
  Nguyen}, \bibinfo{person}{Stephanie Forrest}, {and} \bibinfo{person}{Westley
  Weimer}.} \bibinfo{year}{2012}\natexlab{}.
\newblock \showarticletitle{GenProg: {A} Generic Method for Automatic Software
  Repair}.
\newblock \bibinfo{journal}{\emph{{IEEE} Trans. Software Eng.}}
  \bibinfo{volume}{38}, \bibinfo{number}{1} (\bibinfo{year}{2012}),
  \bibinfo{pages}{54--72}.
\newblock
\urldef\tempurl%
\url{https://doi.org/10.1109/TSE.2011.104}
\showDOI{\tempurl}


\bibitem[\protect\citeauthoryear{Long and Rinard}{Long and Rinard}{2015}]%
        {Long2015}
\bibfield{author}{\bibinfo{person}{Fan Long} {and} \bibinfo{person}{Martin
  Rinard}.} \bibinfo{year}{2015}\natexlab{}.
\newblock \showarticletitle{Staged Program Repair with Condition Synthesis}. In
  \bibinfo{booktitle}{\emph{Proceedings of the 2015 10th Joint Meeting on
  Foundations of Software Engineering}} (Bergamo, Italy)
  \emph{(\bibinfo{series}{ESEC/FSE 2015})}. \bibinfo{publisher}{ACM},
  \bibinfo{address}{New York, NY, USA}, \bibinfo{pages}{166--178}.
\newblock
\showISBNx{978-1-4503-3675-8}
\urldef\tempurl%
\url{https://doi.org/10.1145/2786805.2786811}
\showDOI{\tempurl}


\bibitem[\protect\citeauthoryear{Long and Rinard}{Long and Rinard}{2016}]%
        {Long2016}
\bibfield{author}{\bibinfo{person}{Fan Long} {and} \bibinfo{person}{Martin
  Rinard}.} \bibinfo{year}{2016}\natexlab{}.
\newblock \showarticletitle{Automatic Patch Generation by Learning Correct
  Code}.
\newblock \bibinfo{journal}{\emph{SIGPLAN Not.}} \bibinfo{volume}{51},
  \bibinfo{number}{1} (\bibinfo{date}{Jan.} \bibinfo{year}{2016}),
  \bibinfo{pages}{298--312}.
\newblock
\showISSN{0362-1340}
\urldef\tempurl%
\url{https://doi.org/10.1145/2914770.2837617}
\showDOI{\tempurl}


\bibitem[\protect\citeauthoryear{Miller, Vandome, and McBrewster}{Miller
  et~al\mbox{.}}{2010}]%
        {Miller2010}
\bibfield{author}{\bibinfo{person}{Frederic~P. Miller},
  \bibinfo{person}{Agnes~F. Vandome}, {and} \bibinfo{person}{John McBrewster}.}
  \bibinfo{year}{2010}\natexlab{}.
\newblock \bibinfo{booktitle}{\emph{Apache Maven}}.
\newblock \bibinfo{publisher}{Alpha Press}.
\newblock
\showISBNx{6130652194, 9786130652197}


\bibitem[\protect\citeauthoryear{Monperrus}{Monperrus}{2018}]%
        {Monperrus:survey}
\bibfield{author}{\bibinfo{person}{Martin Monperrus}.}
  \bibinfo{year}{2018}\natexlab{}.
\newblock \showarticletitle{Automatic Software Repair: A Bibliography}.
\newblock \bibinfo{journal}{\emph{ACM Comput. Surv.}} \bibinfo{volume}{51},
  \bibinfo{number}{1}, Article \bibinfo{articleno}{17} (\bibinfo{date}{Jan.}
  \bibinfo{year}{2018}), \bibinfo{numpages}{24}~pages.
\newblock
\showISSN{0360-0300}
\urldef\tempurl%
\url{https://doi.org/10.1145/3105906}
\showDOI{\tempurl}


\bibitem[\protect\citeauthoryear{M\"{u}llerburg}{M\"{u}llerburg}{1983}]%
        {Mullerburg1983}
\bibfield{author}{\bibinfo{person}{Monika A.~F. M\"{u}llerburg}.}
  \bibinfo{year}{1983}\natexlab{}.
\newblock \showarticletitle{The Role of Debugging Within Software Engineering
  Environments}.
\newblock \bibinfo{journal}{\emph{SIGPLAN Not.}} \bibinfo{volume}{18},
  \bibinfo{number}{8} (\bibinfo{date}{March} \bibinfo{year}{1983}),
  \bibinfo{pages}{81--90}.
\newblock
\showISSN{0362-1340}
\urldef\tempurl%
\url{https://doi.org/10.1145/1006142.1006165}
\showDOI{\tempurl}


\bibitem[\protect\citeauthoryear{Pradel and Sen}{Pradel and Sen}{2018}]%
        {Pradel2018}
\bibfield{author}{\bibinfo{person}{Michael Pradel} {and}
  \bibinfo{person}{Koushik Sen}.} \bibinfo{year}{2018}\natexlab{}.
\newblock \showarticletitle{DeepBugs: A Learning Approach to Name-based Bug
  Detection}.
\newblock \bibinfo{journal}{\emph{Proc. ACM Program. Lang.}}
  \bibinfo{volume}{2}, \bibinfo{number}{OOPSLA}, Article
  \bibinfo{articleno}{147} (\bibinfo{date}{Oct.} \bibinfo{year}{2018}),
  \bibinfo{numpages}{25}~pages.
\newblock
\showISSN{2475-1421}
\urldef\tempurl%
\url{https://doi.org/10.1145/3276517}
\showDOI{\tempurl}


\bibitem[\protect\citeauthoryear{Ray, Hellendoorn, Tu, Nguyen, Godhane,
  Bacchelli, and Devanbu}{Ray et~al\mbox{.}}{2016}]%
        {Ray2015}
\bibfield{author}{\bibinfo{person}{Baishakhi Ray}, \bibinfo{person}{Vincent
  Hellendoorn}, \bibinfo{person}{Zhaopeng Tu}, \bibinfo{person}{Connie Nguyen},
  \bibinfo{person}{Saheel Godhane}, \bibinfo{person}{Alberto Bacchelli}, {and}
  \bibinfo{person}{Premkumar Devanbu}.} \bibinfo{year}{2016}\natexlab{}.
\newblock \showarticletitle{On the" Naturalness" of Buggy Code}
  \emph{(\bibinfo{series}{ICSE '16})}. ACM.
\newblock


\bibitem[\protect\citeauthoryear{Sadowski, van Gogh, Jaspan, Soederberg, and
  Winter}{Sadowski et~al\mbox{.}}{2015}]%
        {Sadowski2015}
\bibfield{author}{\bibinfo{person}{Caitlin Sadowski}, \bibinfo{person}{Jeffrey
  van Gogh}, \bibinfo{person}{Ciera Jaspan}, \bibinfo{person}{Emma Soederberg},
  {and} \bibinfo{person}{Collin Winter}.} \bibinfo{year}{2015}\natexlab{}.
\newblock \showarticletitle{Tricorder: Building a Program Analysis Ecosystem}.
  In \bibinfo{booktitle}{\emph{International Conference on Software Engineering
  (ICSE)}}.
\newblock


\bibitem[\protect\citeauthoryear{Saha, Lyu, Lam, Yoshida, and Prasad}{Saha
  et~al\mbox{.}}{2018}]%
        {Saha2018}
\bibfield{author}{\bibinfo{person}{Ripon~K. Saha}, \bibinfo{person}{Yingjun
  Lyu}, \bibinfo{person}{Wing Lam}, \bibinfo{person}{Hiroaki Yoshida}, {and}
  \bibinfo{person}{Mukul~R. Prasad}.} \bibinfo{year}{2018}\natexlab{}.
\newblock \showarticletitle{Bugs.Jar: A Large-scale, Diverse Dataset of
  Real-world Java Bugs}. In \bibinfo{booktitle}{\emph{Proceedings of the 15th
  International Conference on Mining Software Repositories}} (Gothenburg,
  Sweden) \emph{(\bibinfo{series}{MSR '18})}. \bibinfo{publisher}{ACM},
  \bibinfo{address}{New York, NY, USA}, \bibinfo{pages}{10--13}.
\newblock
\showISBNx{978-1-4503-5716-6}
\urldef\tempurl%
\url{https://doi.org/10.1145/3196398.3196473}
\showDOI{\tempurl}


\bibitem[\protect\citeauthoryear{Sliwerski, Zimmermann, and Zeller}{Sliwerski
  et~al\mbox{.}}{2005}]%
        {Sliwerski2005}
\bibfield{author}{\bibinfo{person}{Jacek Sliwerski}, \bibinfo{person}{Thomas
  Zimmermann}, {and} \bibinfo{person}{Andreas Zeller}.}
  \bibinfo{year}{2005}\natexlab{}.
\newblock \showarticletitle{When do changes induce fixes?}. In
  \bibinfo{booktitle}{\emph{International Workshop on Mining Software
  Repositories}}. \bibinfo{publisher}{ACM}.
\newblock
\showISBNx{1-59593-123-6}


\bibitem[\protect\citeauthoryear{Tufano, Watson, Bavota, Di~Penta, White, and
  Poshyvanyk}{Tufano et~al\mbox{.}}{2018a}]%
        {Tufano2018}
\bibfield{author}{\bibinfo{person}{Michele Tufano}, \bibinfo{person}{Cody
  Watson}, \bibinfo{person}{Gabriele Bavota}, \bibinfo{person}{Massimiliano
  Di~Penta}, \bibinfo{person}{Martin White}, {and} \bibinfo{person}{Denys
  Poshyvanyk}.} \bibinfo{year}{2018}\natexlab{a}.
\newblock \showarticletitle{An Empirical Investigation into Learning Bug-fixing
  Patches in the Wild via Neural Machine Translation}. In
  \bibinfo{booktitle}{\emph{Proceedings of the 33rd ACM/IEEE International
  Conference on Automated Software Engineering}} (Montpellier, France)
  \emph{(\bibinfo{series}{ASE 2018})}. \bibinfo{publisher}{ACM},
  \bibinfo{address}{New York, NY, USA}, \bibinfo{pages}{832--837}.
\newblock
\showISBNx{978-1-4503-5937-5}
\urldef\tempurl%
\url{https://doi.org/10.1145/3238147.3240732}
\showDOI{\tempurl}


\bibitem[\protect\citeauthoryear{Tufano, Watson, Bavota, Penta, White, and
  Poshyvanyk}{Tufano et~al\mbox{.}}{2018b}]%
        {Bugs2Fix}
\bibfield{author}{\bibinfo{person}{Michele Tufano}, \bibinfo{person}{Cody
  Watson}, \bibinfo{person}{Gabriele Bavota}, \bibinfo{person}{Massimiliano~Di
  Penta}, \bibinfo{person}{Martin White}, {and} \bibinfo{person}{Denys
  Poshyvanyk}.} \bibinfo{year}{2018}\natexlab{b}.
\newblock \showarticletitle{An Empirical Study on Learning Bug-Fixing Patches
  in the Wild via Neural Machine Translation}.
\newblock \bibinfo{journal}{\emph{CoRR}}  \bibinfo{volume}{abs/1812.08693}
  (\bibinfo{year}{2018}).
\newblock
\showeprint[arxiv]{1812.08693}
\urldef\tempurl%
\url{http://arxiv.org/abs/1812.08693}
\showURL{%
\tempurl}


\bibitem[\protect\citeauthoryear{Williams and Spacco}{Williams and
  Spacco}{2008}]%
        {Williams2008}
\bibfield{author}{\bibinfo{person}{Chadd Williams} {and} \bibinfo{person}{Jaime
  Spacco}.} \bibinfo{year}{2008}\natexlab{}.
\newblock \showarticletitle{{SZZ} Revisited: Verifying when Changes Induce
  Fixes}. In \bibinfo{booktitle}{\emph{Proceedings of the 2008 Workshop on
  Defects in Large Software Systems}} (Seattle, Washington)
  \emph{(\bibinfo{series}{DEFECTS '08})}. \bibinfo{publisher}{ACM},
  \bibinfo{address}{New York, NY, USA}, \bibinfo{pages}{32--36}.
\newblock
\showISBNx{978-1-60558-051-7}
\urldef\tempurl%
\url{https://doi.org/10.1145/1390817.1390826}
\showDOI{\tempurl}


\end{thebibliography}

\end{document}